\documentclass[twocolumn]{revtex4-1}

\usepackage{bm}
\usepackage{comment}
\usepackage{amsmath}
\usepackage{amsfonts}
\usepackage{amssymb}

\date{\today}
\usepackage[dvipdfmx]{graphicx}

\usepackage{color}

\newcommand{\ua}{\uparrow}
\newcommand{\da}{\downarrow}

\newcommand{\la}{\langle}
\newcommand{\ra}{\rangle}
\newcommand{\al}{\alpha}
\newcommand{\sg}{\sigma}

\newcommand{\ep}{\varepsilon}

\begin{document}
\title{
Unconventional Full-Gap Superconductivity in Kondo Lattice with Semi-Metallic Conduction Bands
}

\author{Shoma Iimura$^{1}$} 
\author{Motoaki Hirayama$^2$}
\author{Shintaro Hoshino$^{1}$}

\affiliation{
$^1$Department of Physics, Saitama University, Shimo-Okubo, Saitama 338-8570, Japan\\
$^2$RIKEN Center for Emergent Matter Science (CEMS), Wako, Saitama 351-0198, Japan
}

\begin{abstract}
A mechanism of superconductivity is proposed for the Kondo lattice which has semi-metallic conduction bands with electron and hole Fermi surfaces.
At high temperatures, the $f$ electron's localized spins/pseudospins are fluctuating between electron and hole Fermi surfaces to seek for a partner to couple with. 
This system tries to resolve this frustration at low temperatures and chooses to construct a quantum mechanically entangled state composed of the Kondo singlet with electron surface and that with hole surface, to break the U(1) gauge symmetry.
The corresponding order parameter is given by a composite pairing amplitude as a three-body bound states of localized spin/pseudospin, electron and hole.
The electromagnetic response is considered, where composite pair itself does not contribute to the Meissner effect, but the induced pair between conduction electrons, which inevitably mixes due to e.g. a band cutoff effect at high energies, carries the superconducting current under the external field.
Possible applications to real heavy-electron materials are also discussed.
\end{abstract}

\maketitle

The mechanism of unconventional superconductivities is an important issue in condensed matter physics \cite{Steglich79,Jerome80,Bednorz86} in designing a guiding principle to find new superconductors. The superconducting materials range over a broad class of correlated electron materials including cuprates, iron pnictides and organic compounds \cite{Norman11}. Among them, the heavy electron materials with nearly localized $f$ electrons are typical systems showing unconventional superconductivity, and the identification of the mechanisms has still remained open question since the discovery of CeCu$_2$Si$_2$ \cite{Steglich79} and UBe$_{13}$ \cite{Ott83} in lanthanide- and actinide-based materials. Whereas the recent advanced experiments allow to determine the symmetry or structure of pairing gap functions, the underlying mechanisms have not yet been fully clarified.

\begin{figure}[b]
\begin{center}
\includegraphics[width=85mm]{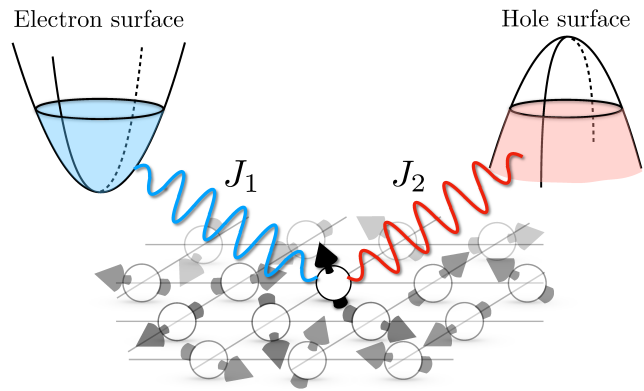}
\caption{(color online). Schematic picture for pairing in Kondo lattice with semi-metallic conduction bands.}
\label{fig:concept}
\end{center}
\end{figure}

Recently, specific heat measurements in a rotational magnetic field have revealed the full-gap nature of the superconducting states in CeCu$_2$Si$_2$ \cite{Kittaka14} and UBe$_{13}$ \cite{Shimizu15}, in addition to a well known cerium-based $s$-wave superconductor CeRu$_2$ \cite{Hedo98}.
This is in contrast with the conventional notion that the strongly correlated electrons favor a spatially non-local and anisotropic pairing.
Hence, it is desirable to identify a new mechanism for full-gap superconductivity that is specific to heavy-electron materials. 
To this end, we here focus on the physics arising from a semi-metallic band structure having both electron and hole Fermi surfaces.
One of the characteristics of the correlated semi-metal is an emergence of excitonic insulator \cite{Mott61,Halperin68,Kunes15}, where the Coulomb interactions reconstruct the electronic states near the Fermi level and form a gap at the Fermi level.
Another interesting problem, which is discussed in this paper, is to consider the situation where the semi-metal interacts quantum-mechanically with magnetic ions through the antiferromagnetic Kondo coupling as illustrated in Fig.~\ref{fig:concept}.

We show in this paper that the Kondo lattice with semi-metallic conduction bands exhibits a full-gap superconductivity as a consequence of resolution of a frustration associated with the multichannel Kondo effect.
Underlying physical picture is the following:
at high temperatures electron and hole bands are competing in screening the localized moments (see Fig.~\ref{fig:concept}), and at low temperatures they decide to mix the two kinds of Kondo singlet states quantum-mechanically.
As a result, the electrons and holes are superposed and superconductivity occurs since the U(1) gauge symmetry is broken.
The resultant order parameter is the composite pair amplitude \cite{Emery92,Balatsky93,Coleman93,Coleman95}, which has been proposed in the multi-channel Kondo/Anderson lattices \cite{Emery93,Coleman99,Jarrell97,Anders02,Anders02-2,Flint08,Flint10,Flint14,Hoshino14}.
A new perspective proposed here is that the stability of this composite pairing is closely related to the semi-metallic conduction bands. 
The mechanism is also related closely to the concept of frustration.
It has been argued that the frustration generates intriguing quantum states out of classically degenerate states.
The well-known example is the spin liquid caused by the geometrical frustration in the magnetically interacting systems \cite{Zhou17,Savary17}.
In our setup, the frustration arises in the form of the competing screening channels in Kondo systems, which is resolved by utilizing the quantum states, and even realizes the gauge-symmetry broken state. 

For theoretical analysis, we take the parabolic dispersion for conduction electrons, which makes the physical picture clear with a {\it pure} composite pairing state without conventional one-body pair amplitudes.
The electromagnetic response is also considered based on the microscopic Hamiltonian, and the pure composite pair does not directly couple to electromagnetic field in linear response.
Instead, with the consideration of lattice regularization (or band cutoff effect) that modifies the parabolic dispersions far away from Fermi surfaces, a usual one-body pair amplitude is secondarily induced and then show electromagnetic linear-response.

We use the following Hamiltonian for the Kondo lattice (KL) with semi-metallic conduction bands (SMCB):
\begin{align}
\mathcal H
&= 
\sum_{\alpha=1,2} \sum_{\sigma=\ua,\da}
\int_{\bm k\in \mathcal K_\al}
\frac{d{\bm k}}{(2\pi)^3}
\psi_{\bm k\alpha \sigma}^\dagger
\xi_{\bm{k}\alpha}
\psi_{\bm k\alpha\sigma} \nonumber \\
&\ \ +
\frac{1}{2}\sum_{\al\sigma \sigma'} J_\alpha \int d{\bm r}~{\bm S}({\bm r}) \cdot \psi_{\alpha\sigma}^\dagger ({\bm r}) {\bm{\sigma}}_{\sigma \sigma'} \psi_{\alpha \sigma'}^{ }({\bm r}),
\label{SMKL_continuous}
\end{align}
where 
$\psi_{\alpha \sigma}^{ } ({\bm r})$ ($\psi_{\alpha \sigma}^\dagger ({\bm r})$) is the annihilation (creation) operator of electrons at ${\bm r}$ in conduction band $\alpha$ with spin $\sigma$ and $\xi_{\bm{k}\alpha}$ is the one-body energy of the conducting electron, $\xi_{\bm{k}\alpha} = \hbar^2 (\bm{k}-\bm{K}_\alpha)^2/(2m_\alpha) - \mu_\alpha$.
$m_\alpha$ denotes an effective mass of the conduction electrons: $m_1>0$ (electron) and $m_2<0$ (hole).
The chemical potential $\mu_\al$ is introduced for each bands.
The condition $m_1 \mu_1 = m_2 \mu_2$ is satisfied in compensated metals.
$\bm{K}_\alpha$ denotes the centers of electron and hole pockets, and the wave vector summation is taken over the range $\left| \bm{k} -\bm{K}_\al \right| < k_c $ (denoted by $\mathcal K_\al$) where $k_c$ is a wavevector cutoff.
$J_\alpha$ is the Kondo coupling. The localized spin operator is given by $\bm S(\bm r) = \sum_{i} \bm S_i \delta(\bm r-\bm R_i)$ where the $f$-electron spins $\bm S_i$ are localized at the lattice sites $\bm R_i$. ${\bm{\sigma}}$ is a Pauli matrix.
Schematic illustration of our model is shown in Fig.~\ref{fig:concept}.

In the above, we have assumed the Kramers doublet for the $f$-electron degrees of freedom. On the other hand, for the non-Kramers doublet realized in $f^2$ configuration of U and Pr ions, which is not necessarily associated with time-reversal symmetry, the localized state is described by a pseudospin $\bm T$ \cite{Cox87,Cox98}.
This pseudospin interacts with semi-metallic conduction bands through the nonmagnetic degrees of freedom $\al=1,2$.
The simplest interaction takes the form
\begin{align}
\label{non_Kramers}
\mathcal H_{\rm int} &= 
\frac{1}{2}\sum_{\al\al'\sigma} J \int d{\bm r}~{\bm T}({\bm r}) \cdot \psi_{\alpha\sigma}^\dagger ({\bm r}) {\bm{\sigma}}_{\al\al'} \psi_{\al' \sigma}^{ }({\bm r}).
\end{align}
In this case, the equivalence between $\sg=\ua$ and $\da$ for conduction electrons is protected by the time-reversal symmetry 
and the competition arises in forming a Kondo singlet state.
Since the mean-field solution is same as the Kramers case [see Supplementary Materials (SM) A], in the following we mainly focus on the Kramers system described by Eq.~\eqref{SMKL_continuous} 
to simplify the discussion. 
Now we apply the mean-field theory which effectively describe the superconducting state resulting from the superposition of ``electron'' and ``hole'' through the multichannel Kondo effect. We first write the localized spin in terms of pseudofermion as
$\label{local_spin_fermionic}
{\bm S}({\bm r}) = \frac{1}{2} \sum_{\sigma \sigma'} f_\sigma^\dagger ({\bm r}) {\bm{\sigma}}_{\sigma \sigma'} f_{\sigma'}^{ } ({\bm r}),
$
with the local constraint
$\sum_\sigma \la f_{\sigma}^\dagger ({\bm r}) f_{\sigma}^{ } ({\bm r}) \ra = n_f$,
where $f_{\sigma} ({\bm r})$ ($f_{\sigma}^\dagger ({\bm r})$) is the annihilation (creation) operator of the $f$-electron and $n_f$ is an $f$-electron density.
The mean-fields are introduced as
\begin{align}
&\label{V}
V = \frac{3}{4} J_1 \langle f_\sigma^{}({\bm r}) \psi_{1\sigma}^\dagger ({\bm r})\rangle,\\
\label{W}
&W \epsilon_{\sigma\sigma'} = \frac{3}{4} J_2 \langle f_\sigma^{ }({\bm r}) \psi_{2\sigma'}^{ } ({\bm r}) \rangle,
\end{align}
where $\hat{\epsilon} = i \hat{\sigma}^y$ is the antisymmetric unit tensor.
The mean-fields $V$ and $W$ respectively represent ``hybridization'' and ``pair-condensation'' between the pseudofermion and the conduction electrons \cite{Coleman99, Flint08, Hoshino14-2, Kusunose16}. Although the mean-fields $V$ and $W$ are introduced asymmetrically with respect to the index $\al=1,2$, this situation is related to the fact that the way of introduction of pseudofermions is not unique \cite{Hoshino14-2}. Hence, if we consider the physical quantities in terms of the original $\bm S$, the symmetries between $\al=1,2$ and between $\sg=\ua,\da$ are preserved. 
The nonzero self-consistent solution is obtained when the Kondo coupling is antiferromagnetic (see Eq.(\ref{self_consistent_equation})). 
Our pseudofermion approach is justified in the low-temperature limit where the collective Kondo effects are fully activated as in the ordinary Kondo lattice  \cite{Zhang00, Capponi01}.
In this regime, the resonant $f$-electron level is generated near the Fermi surfaces.

\begin{figure}[t]
\begin{center}
\includegraphics[width=75mm]{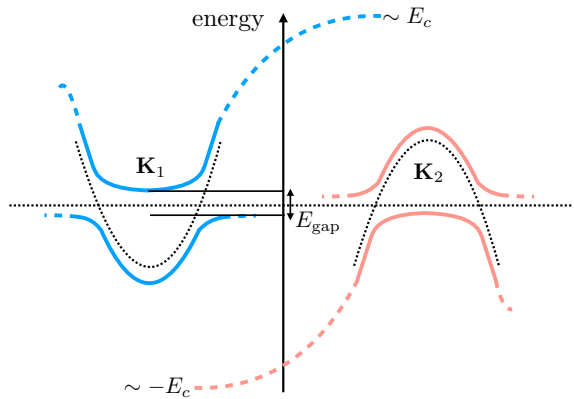}
\caption{(color online). Illustration for the band structures in superconducting state.
The energy cutoff $E_c$ corresponding to the band width is needed for the electromagnetic response.
}
\label{fig:band}
\end{center}
\end{figure}

We now show the energy dispersion relation in superconducting state, which is obtained as
\begin{align}
E_{{\bm k}\alpha \pm} &= \frac{1}{2} \left( \xi_{\bm{k}\alpha} \pm \sqrt{ \xi_{\bm{k}\alpha}^2 + 4 \left| V_\alpha \right|^2} \right),
\label{eq:dispersion}
\end{align}
for each $\bm k \in \mathcal K_\al$. We have defined $V_\al$ by $V_1 = V$, and $V_2 = W$.
The solid lines in Fig.~\ref{fig:band} show the dispersion relation described by Eq.~\eqref{eq:dispersion}. The single-particle spectrum has the fully opened energy gap at the Fermi level. In addition, unlike the $s$-wave superconductor in the Bardeen-Cooper-Schrieffer (BCS) theory, the minimal energy gap is indirect with the amplitude $E_{\mathrm{gap},\al} \sim \left| V_\alpha \right|^2/|\mu_\al| + \left| V_\al \right|^2/D_\al$ where 
$D_\al$ ($\gtrsim |\mu_\al|$) denotes the energy range in which the conduction electrons are involved into the condensation. This energy gap is reflected in the thermodynamic properties such as a temperature dependence of the specific heat. On the other hand, the direct energy gap shows minimum at the Fermi momentum with its amplitude being $2 \left| V_\al \right|$ ($\gg E_{\rm gap,\al}$), and is reflected in the optical conductivity.

Although $V$ and $W$ can be regarded as the order parameters in the effective model, the pseudofermions introduced to describe the low-energy excitations are not real but virtual physical degrees of freedom. Then we need to seek for the other pair amplitude that remains in the original model without considering pseudofermions. 
For a U(1) symmetry broken state, one naively expects a finite pair amplitude $\la \psi_{\bm k \al \sg} \psi_{-\bm k, \al' \sg'} \ra$ for the conduction electrons. However, with the present effective low-energy Hamiltonian, any pair amplitude composed of conduction electrons becomes zero. This is due to the separation of the regions $\mathcal{K}_1$ and $\mathcal{K}_2$ in our model. Instead, the composite pair amplitude is the appropriate order parameter, which is given by
\begin{align}
&\Phi_{\mu,\sigma\sigma'}^{\alpha \alpha'} (\bm{R};\bm{r},\bm{r}') =\langle S^\mu (\bm{R}) \psi_{\alpha \sigma}^{ } (\bm{r}) \psi_{\alpha'\sigma'}^{ } (\bm{r}') \rangle \nonumber \\
\label{eq:composite_pair}
&=
\frac{1}{2} \left( \hat{\sigma}^\mu \hat{\epsilon}\right)_{\sigma \sigma'} \epsilon_{\al \al'} V^* W  F_{\al} (\bm{r}-\bm{R} ) F_{\al'}^{*} ( \bm{r}' - \bm{R} ),
\end{align}
where the function $F_\al$ behaves as
\begin{align}
\label{eq:correlation_1}
&F_\al (\bm r =\bm 0) = \rho_\alpha (0) \mathrm{log}\left( \frac{D_\al |\mu_\al| }{\left| V_\alpha \right|^2} \right),
\\
\label{eq:correlation_2}
&F_\alpha (|\bm r|\to \infty) = 2\mathrm{e}^{i {\bm K}_\al \cdot {\bm r}} \rho_\alpha (0)\frac{\mathrm{sin}\left( k_{\mathrm{F}\alpha}r \right)\mathrm{e}^{-r/{\xi_\alpha} }}{k_{\mathrm{F}\alpha}r}.
\end{align}
See SM B for derivation.
Here $\rho_\alpha (0)$ is the density of states of the conduction electron at the Fermi level. $k_{\mathrm{F}\alpha} = \sqrt{2m_\alpha \mu_\alpha}/\hbar$ stands for the Fermi momentum for each band. $\xi_\alpha$ represents the coherence length defined by $\xi_\alpha = \left| \mu_\alpha \right|/k_{\mathrm{F}\alpha} \left| V_\alpha \right|$, which is determined by the energy scale of $\left| V_\al \right|$. Since $E_{\mathrm{gap},\al} \ll \left| V_\al \right|$, the coherence length becomes much shorter than that expected by the amplitude of the energy gap. This is due to the nearly localized nature of the composite pairs involving the $f$-electron.

The composite pair amplitude $\Phi (\bm{R},\bm{r},\bm{r}')$ depends on the relative coordinates $\bm{r} - \bm{R}$ and $\bm{r}'-\bm{R}$ measured from the position of the localized spin, whereas it does not depend on that between conduction electrons, $\bm{r}-\bm{r}'$. 
This nature reflects the superposition between the electron-Kondo singlet and the hole-Kondo singlet and not a simple pair-condensation of conduction electrons.
We have thus demonstrated that the composite pair amplitude is generated in the SMCB-KL.
We also comment on the relation between composite pair and odd-frequency pairing.
The composite order parameter has been discussed in the context of the odd-frequency superconductor to date \cite{Emery92, Balatsky93, Coleman93}. 
However, our result shows that, as long as the low-energy property is concerned, there is no pair amplitude between conduction electrons.
In this sense, the composite order parameter does not directly mean the presence of odd-frequency pairing.
This point can be clarified with our effective model in Eq.~\eqref{SMKL_continuous}.

The values of mean-fields can be determined by solving the self-consistent equations given as follows:
\begin{align}
\label{self_consistent_equation}
1 &= \frac{3}{4} \frac{J_\alpha}{\Omega} \sum_{\bm{k}\in \mathcal K_\alpha} \frac{ f(E_{\bm{k}\alpha -}) - f(E_{\bm{k}\alpha +})}{E_{\bm{k}\alpha+} - E_{\bm{k}\alpha-}},
\end{align}
where $f(x) = 1/(\mathrm{e}^{\beta x} +1)$ is the Fermi-Dirac distribution function and $\Omega$ is a volume of the system. 
The critical temperature $T_{c,\al}$ can be obtained for each mean fields from the above self-consistent equation as
\begin{align}
\label{eq:Tc_al}
k_{\rm B} T_{c,\alpha} = \frac{2 \mathrm{e}^\gamma}{\pi} \sqrt{D_\al \left| \mu_\al \right|}~\mathrm{exp}\left[ - \frac{4}{3J_\alpha \rho_\alpha (0)}\right],
\end{align}
where $\gamma$ is the Euler constant and the prefactor is $2\mathrm{e}^\gamma/\pi \simeq 1.13$.
There is also the relation $k_{\rm B} T_{c,\al}\propto E_{\mathrm{gap},\al}$, where the coefficient is order of unity but is still dependent on the parameters such as $D_\al$ and $\mu_\al$ due to the magnitude relation $D_\al \sim |\mu_\al|$. See SM C for more details.
Since only the Kondo effect is involved in our theory, the appearance of the Kondo gap is natural as the characteristic energy scale, but it is much reduced from that in the usual heavy electrons because of a small density of states in semimetals. 
For a rough estimation, we use the expression $T_K = D \exp(-1/\lambda)$ where $D$ is a half bandwidth and $\lambda$ is a density of states at Fermi level multiplied with the Kondo coupling.
If we take the bandwidth $2D=10^4$K and $\lambda = 0.3$, we get $T_K \simeq 180$K. For SMCB, $\lambda$ is smaller and we get $T_K\simeq 0.2$K when we choose $\lambda = 0.1$. 
We also examine the thermodynamic stability of the present ordered state in the low-temperature limit, and obtain the following free energy density 
$F(T) \simeq F_0(0) - 2 \sum_\alpha \rho_\alpha (0) \left| V_\alpha \right|^2$, where $F_0(T)$ denotes the component in the normal state.
Therefore, the composite pair amplitude contributes to the energy-lowering in the low-temperature limit.

The energy scales of the critical temperature should be considered separately for the Kramers case and the non-Kramers case. 
In general, the electron and hole bands can be asymmetric, and then $T_{c,1}$ is different from $T_{c,2}$
in the Kramers case. 
Since the self-consistent equations for $T_{c,1}$ and $T_{c,2}$ are independent in the low-energy effective model, the superconducting critical temperature is estimated by $T_c \sim \min(T_{c,1},T_{c,2})$. The larger value of $T_{c,\al}$ is regarded as a crossover scale for the single-channel Kondo effect. 
For a non-Kramers doublet system, on the other hand, the equivalence of the channels is protected by the time-reversal symmetry, thus only one energy scale appears for $T_c$.

So far, we have considered only the degrees of freedom near the Fermi level, and concluded the presence of composite pair amplitude in Eq.~\eqref{eq:composite_pair} and the absence of the conventional Cooper pairs.
However, with this situation, the composite pairs are not carried by the external field which acts only on the conduction electrons.
The Meissner effect is then absent. Hence we need to consider the secondarily induced pair amplitude among the conduction electrons for the electromagnetic response.
The property of this induced pair amplitude depends on the detail of the high-energy region and is specific to the material details.
As one of the origins, we consider the finite cut-off of the energy dispersion at high energy $E_c$ ($\sim D_1,D_2$)  (see Fig.~\ref{fig:band}). 
We note that the regions $\mathcal{K}_1$ and $\mathcal{K}_2$ now overlap with each other.
This is regarded as a lattice-regularization for the energy dispersion.
The contribution from the high-energy region can be included perturbatively by expanding the physical quantities as a series of the orders of $E_c^{-1}$. (See SM D for derivation.) 
Thus the pair amplitude composed of conduction electrons can be finite in general for a tight-binding lattice, which is obtained as
\begin{align}
\la \psi_{\bm k\al\sg} \psi_{-\bm k,\al' \sg'}\ra =- \frac{V^* W}{E_c}\epsilon_{\sg \sg'} \sg^x_{\al \al'}
\sum_{\al''}
\frac{\theta \left(k_c - \left| {\bm k}-{\bm K}_{\al''}
\right| \right)}{\sqrt{\xi_{{\bm k}\al''}^2 + 4 \left| V_{\al''} \right|^2}} .
\end{align}
The coherence length $\xi_c$ for this induced Cooper pair is $\xi_c = \max \{\xi_1,\xi_2\}$ which is the same order of magnitude as that for the composite pairs in Eq.~\eqref{eq:composite_pair}.

Electromagnetic response functions can be calculated based on the Kubo formula \cite{Kubo57}. 
Within the linear response of the vector potential $\bm{A}(\bm{q})$, which is applied on the conduction electrons, we can write the total current as $j_{\text{tot}}^\mu = - \sum_\nu K_{\mu \nu} A_\nu$. 
The introduced kernel is separated into two terms as $K_{\mu\nu} = K_{\mu\nu}^d + K_{\mu\nu}^p$, which describes diamagnetic and paramagnetic contributions, respectively. 
Each term is given as
\begin{align}
K_{\mu\nu}^p (\bm{q}) &= - \int_0^\beta d\tau \langle \hat{j}_\mu^p (\bm{q},\tau) \hat{j}_\nu^p (-\bm{q}) \rangle, \\
K_{\mu\nu}^d (\bm{q}) &= \delta_{\mu\nu} \frac{e^2}{\hbar^2} 
\sum_{\alpha,\sg} 
\int \frac{d\bm{k}}{(2\pi)^3}
\frac{\partial^2 
\varepsilon_{\bm{k}\alpha}}{\partial k_\mu^2} \langle \psi_{\bm{k}\alpha\sg}^\dagger \psi_{\bm{k}\alpha\sg}^{ } \rangle.
\end{align}
The paramagnetic current operator $\hat{j}_\mu^p$ is defined as
\begin{align}
\hat{\bm j}^p
 (\bm{q}) = e\sum_{\alpha,\sg} \int d\bm{k}~\psi_{\bm{k}-\bm{q}/2,\alpha\sg}^\dagger 
\bm v_{\bm k\al}
\psi_{\bm{k}+\bm{q}/2,\alpha\sg}^{ },
\end{align}
where $e<0$ denotes the charge of the electron and $\bm v_{\bm k\al} = \hbar^{-1}\partial_{\bm k} \ep_{\bm k\al}$ is the velocity.
We note that the energy $\varepsilon_{\bm k\al}$ represents the tight-binding band and includes the high-energy part shown in Fig.~\ref{fig:band}.

\begin{figure}[t]
\begin{center}
\includegraphics[width=80mm]{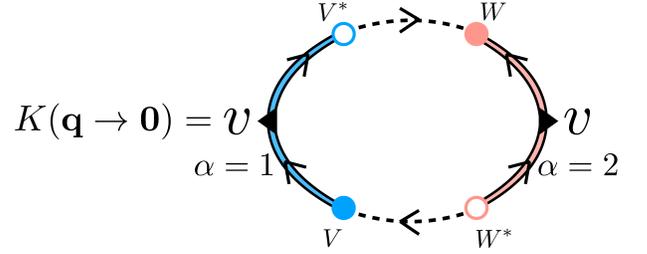}
\caption{(color online). Illustration for the lowest-order diagram that contributes to the Meissner kernel. The blue (pink) solid and the black dashed lines with the arrow represent the conduction electron (hole) and the pseudofermion, respectively.
}
\label{fig:kernel}
\end{center}
\end{figure}

We neglect the vertex correction for the current correlation function as in the dynamical mean-field theory which is exact in infinite dimensions and is regarded as a good approximation for three dimensions \cite{Georges96,Khurana90}.
Then the 
leading-order contribution to the response function is diagramatically shown in Fig.~\ref{fig:kernel}.
At zero temperature, the electromagnetic kernel is evaluated as
\begin{align}
\label{eq:Meissner_kernel}
K_{\mu\nu} (\bm q\to \bm 0) &\simeq \sum_\alpha \frac{4 \left| V_1 \right|^2 \left| V_2 \right|^2}{\left| V_\alpha \right|^2 E_c^2} \frac{n_\alpha e^2}{\left| \tilde{m}_\alpha \right|} \delta_{\mu\nu},
\end{align}
where $n_\alpha$ is the particle number of the electron $(\alpha =1)$ and the hole $(\alpha =2)$. $\tilde{m}_\alpha (\sim -m_\alpha )$ denotes the effective mass $\varepsilon_{\bm{k}\bar{\alpha}}$ with $\bm{k}\in \mathcal K_\al$ (here $\bar \al$ is the complementary component of $\al$ such as $\bar 1=2$). See SM E for the detailed derivation. 
The magnitude of the kernel is smaller than the usual BCS superconductors by the factor of $(|V_\al|/E_c)^2$.
Hence, the superconducting state has a large magnetic penetration depth reflecting the small Cooper pair amplitude $\la \psi_{\bm k\al\sg} \psi_{-\bm k,\al' \sg'}\ra$.
This is because both $K^p$ and $K^d$ have the finite and almost the same amplitudes at zero temperature.
Therefore, the cancelation between them occurs, and the situation is similar to that for the normal Kondo insulator.
In the presence of the induced pair amplitude, however, the paramagnetic (diamagnetic) term slightly decreases (increases).
The remaining contribution, which is the order of $E_c^{-2}$, plays a role of the superconducting current (See SM E).
In this way, although the induced pair amplitude does not contribute to the condensation energy as we mentioned in the previous section, it plays an essential role for the electromagnetic response expected in the superconducting state.

In this paper we have established the mechanism for the full-gap superconductivity characteristic for 
the semi-metals along with the localized spin/pseudospins. 
We here comment on the consideration of some competing orders.  In the correlated semi-metals, excitonic insulator states \cite{Mott61,Halperin68,Kunes15} or metallic CDWs \cite{Mattheiss73} would compete with the present mechanism of superconductivity.
While such instabilities arise only when the conduction bands have a nesting of the Fermi surface, the present mechanism works when there are the semi-metallic conduction bands regardless of the details of Fermi surfaces. 

Finally, let us discuss the candidate materials of the present superconducting state. We propose the two relevant materials $\mathrm{U} \mathrm{Be}_{13}$ \cite{Ott83} and $\mathrm{Pr} \mathrm{Ir}_2 \mathrm{Zn}_{20}$ \cite{Onimaru11}. For $\mathrm{U}\mathrm{Be}_{13}$, the system undergoes a phase transition into unconventional superconductivity from a non-Fermi liquid state, and the multi-channel Kondo effect has been suspected as a cause \cite{Cox87,Cox98}. 
We speculate that $\rm{U}\rm{Be}_{13}$ can be a candidate material for our scenario based on the following two reasonable assumptions. First, 
as seen from its chemical composition, the number of $\mathrm{Be}$ inside the unit-cell is much larger than the magnetic uranium ion.
Therefore, the presence of the uranium atom does not affect much the conduction band structure. Second, 
the electronic structure of $\mathrm{Be}$ in the atomic limit is partially filled L-shell, which is composed of fully filled $2s$-orbital and empty $2p$-orbital. 
In a crystal environment, the $s$-band and the $p$-band resulting from the inter-berylium hopping bury the atomic gap and the semi-metallic conduction bands are formed.
Indeed, the electronic structures of the elemental substance of Be has a semi-metallic character \cite{Loucks64,Inoue73}. 
Hence, $\mathrm{U}\mathrm{Be}_{13}$ can be regarded as the embedded magnetic $\mathrm{U}$ atom in the sea of the electrons which have electron and hole conduction bands. Thus the SMCB-KL considered in this paper can be realized in this material. 

As for Pr-based materials, $\mathrm{Pr} \mathrm{Ir}_2 \mathrm{Zn}_{20}$, shows a peculiar superconductivity with the multi-channel Kondo behavior \cite{Onimaru11,Yamane18}, can also be regarded as the promising candidate material for the SMCB-KL, since the $\mathrm{Zn}$ has the fully-filled $3d$ and $4s$ orbitals and an energy gap is needed to locate an additional electron in higher-energy $4p$ orbital. 
Indeed the elemental substance of $\mathrm{Zn}$ shows semi-metallic character \cite{Stark67,Allen68}. 
In addition, the number of $\mathrm{Zn}$ is much larger than that of $\mathrm{Pr}$. 
Hence the situation is close to $\mathrm{U} \mathrm{Be}_{13}$, and the formation of the SMCB-KL is expected. 
Further studies such as realistic band calculations combined with electronic correlation effects are needed to establish the actual realization of our proposal in real materials, which are left as interesting future issues. 
In addition to these compounds, the idea of the SMCB-KL can be applied to a wider class of 
materials with the semi-metallic conduction bands plus localized moments, 
and gives a general guiding principle to find unconventional superconductors.

\vspace{2mm}
{\bf Acknowledgement:}
This work was supported by the Japan Society for Promotion of Science (JSPS) KAKENHI Grants No.~18K13490, No.~16H04021, and No.~18H01176.

%


\begin{thebibliography}{42}%
\makeatletter
\providecommand \@ifxundefined [1]{%
 \@ifx{#1\undefined}
}%
\providecommand \@ifnum [1]{%
 \ifnum #1\expandafter \@firstoftwo
 \else \expandafter \@secondoftwo
 \fi
}%
\providecommand \@ifx [1]{%
 \ifx #1\expandafter \@firstoftwo
 \else \expandafter \@secondoftwo
 \fi
}%
\providecommand \natexlab [1]{#1}%
\providecommand \enquote  [1]{``#1''}%
\providecommand \bibnamefont  [1]{#1}%
\providecommand \bibfnamefont [1]{#1}%
\providecommand \citenamefont [1]{#1}%
\providecommand \href@noop [0]{\@secondoftwo}%
\providecommand \href [0]{\begingroup \@sanitize@url \@href}%
\providecommand \@href[1]{\@@startlink{#1}\@@href}%
\providecommand \@@href[1]{\endgroup#1\@@endlink}%
\providecommand \@sanitize@url [0]{\catcode `\\12\catcode `\$12\catcode
  `\&12\catcode `\#12\catcode `\^12\catcode `\_12\catcode `\%12\relax}%
\providecommand \@@startlink[1]{}%
\providecommand \@@endlink[0]{}%
\providecommand \url  [0]{\begingroup\@sanitize@url \@url }%
\providecommand \@url [1]{\endgroup\@href {#1}{\urlprefix }}%
\providecommand \urlprefix  [0]{URL }%
\providecommand \Eprint [0]{\href }%
\providecommand \doibase [0]{http://dx.doi.org/}%
\providecommand \selectlanguage [0]{\@gobble}%
\providecommand \bibinfo  [0]{\@secondoftwo}%
\providecommand \bibfield  [0]{\@secondoftwo}%
\providecommand \translation [1]{[#1]}%
\providecommand \BibitemOpen [0]{}%
\providecommand \bibitemStop [0]{}%
\providecommand \bibitemNoStop [0]{.\EOS\space}%
\providecommand \EOS [0]{\spacefactor3000\relax}%
\providecommand \BibitemShut  [1]{\csname bibitem#1\endcsname}%
\let\auto@bib@innerbib\@empty
\bibitem [{\citenamefont {Steglich}\ \emph {et~al.}(1979)\citenamefont
  {Steglich}, \citenamefont {Aarts}, \citenamefont {Bredl}, \citenamefont
  {Lieke}, \citenamefont {Meschede}, \citenamefont {Franz},\ and\ \citenamefont
  {Sch\"afer}}]{Steglich79}%
  \BibitemOpen
  \bibfield  {author} {\bibinfo {author} {\bibfnamefont {F.}~\bibnamefont
  {Steglich}}, \bibinfo {author} {\bibfnamefont {J.}~\bibnamefont {Aarts}},
  \bibinfo {author} {\bibfnamefont {C.~D.}\ \bibnamefont {Bredl}}, \bibinfo
  {author} {\bibfnamefont {W.}~\bibnamefont {Lieke}}, \bibinfo {author}
  {\bibfnamefont {D.}~\bibnamefont {Meschede}}, \bibinfo {author}
  {\bibfnamefont {W.}~\bibnamefont {Franz}}, \ and\ \bibinfo {author}
  {\bibfnamefont {H.}~\bibnamefont {Sch\"afer}},\ }\href {\doibase
  10.1103/PhysRevLett.43.1892} {\bibfield  {journal} {\bibinfo  {journal}
  {Phys. Rev. Lett.}\ }\textbf {\bibinfo {volume} {43}},\ \bibinfo {pages}
  {1892} (\bibinfo {year} {1979})}\BibitemShut {NoStop}%
\bibitem [{\citenamefont {J\"erome}\ \emph {et~al.}(1980)\citenamefont
  {J\"erome}, \citenamefont {Mazaud}, \citenamefont {Ribault},\ and\
  \citenamefont {Bachgaard}}]{Jerome80}%
  \BibitemOpen
  \bibfield  {author} {\bibinfo {author} {\bibfnamefont {D.}~\bibnamefont
  {J\"erome}}, \bibinfo {author} {\bibfnamefont {A.}~\bibnamefont {Mazaud}},
  \bibinfo {author} {\bibfnamefont {M.}~\bibnamefont {Ribault}}, \ and\
  \bibinfo {author} {\bibfnamefont {K.}~\bibnamefont {Bachgaard}},\ }\href@noop
  {} {\bibfield  {journal} {\bibinfo  {journal} {J. Phys. Lett.}\ }\textbf
  {\bibinfo {volume} {41}},\ \bibinfo {pages} {95} (\bibinfo {year}
  {1980})}\BibitemShut {NoStop}%
\bibitem [{\citenamefont {Bednorz}\ and\ \citenamefont
  {M\"uller}(1986)}]{Bednorz86}%
  \BibitemOpen
  \bibfield  {author} {\bibinfo {author} {\bibfnamefont {J.~G.}\ \bibnamefont
  {Bednorz}}\ and\ \bibinfo {author} {\bibfnamefont {K.~A.}\ \bibnamefont
  {M\"uller}},\ }\href@noop {} {\bibfield  {journal} {\bibinfo  {journal} {Z.
  Phys. B}\ }\textbf {\bibinfo {volume} {64}},\ \bibinfo {pages} {189}
  (\bibinfo {year} {1986})}\BibitemShut {NoStop}%
\bibitem [{\citenamefont {{For a review, see M.~R. Norman}}(2011)}]{Norman11}%
  \BibitemOpen
  \bibfield  {author} {\bibinfo {author} {\bibnamefont {{For a review, see
  M.~R. Norman}}},\ }\href {\doibase 10.1126/science.1200181} {\bibfield
  {journal} {\bibinfo  {journal} {Science}\ }\textbf {\bibinfo {volume}
  {332}},\ \bibinfo {pages} {196} (\bibinfo {year} {2011})}\BibitemShut
  {NoStop}%
\bibitem [{\citenamefont {Ott}\ \emph {et~al.}(1983)\citenamefont {Ott},
  \citenamefont {Rudigier}, \citenamefont {Fisk},\ and\ \citenamefont
  {Smith}}]{Ott83}%
  \BibitemOpen
  \bibfield  {author} {\bibinfo {author} {\bibfnamefont {H.~R.}\ \bibnamefont
  {Ott}}, \bibinfo {author} {\bibfnamefont {H.}~\bibnamefont {Rudigier}},
  \bibinfo {author} {\bibfnamefont {Z.}~\bibnamefont {Fisk}}, \ and\ \bibinfo
  {author} {\bibfnamefont {J.~L.}\ \bibnamefont {Smith}},\ }\href {\doibase
  10.1103/PhysRevLett.50.1595} {\bibfield  {journal} {\bibinfo  {journal}
  {Phys. Rev. Lett.}\ }\textbf {\bibinfo {volume} {50}},\ \bibinfo {pages}
  {1595} (\bibinfo {year} {1983})}\BibitemShut {NoStop}%
\bibitem [{\citenamefont {Kittaka}\ \emph {et~al.}(2014)\citenamefont
  {Kittaka}, \citenamefont {Aoki}, \citenamefont {Shimura}, \citenamefont
  {Sakakibara}, \citenamefont {Seiro}, \citenamefont {Geibel}, \citenamefont
  {Steglich}, \citenamefont {Ikeda},\ and\ \citenamefont
  {Machida}}]{Kittaka14}%
  \BibitemOpen
  \bibfield  {author} {\bibinfo {author} {\bibfnamefont {S.}~\bibnamefont
  {Kittaka}}, \bibinfo {author} {\bibfnamefont {Y.}~\bibnamefont {Aoki}},
  \bibinfo {author} {\bibfnamefont {Y.}~\bibnamefont {Shimura}}, \bibinfo
  {author} {\bibfnamefont {T.}~\bibnamefont {Sakakibara}}, \bibinfo {author}
  {\bibfnamefont {S.}~\bibnamefont {Seiro}}, \bibinfo {author} {\bibfnamefont
  {C.}~\bibnamefont {Geibel}}, \bibinfo {author} {\bibfnamefont
  {F.}~\bibnamefont {Steglich}}, \bibinfo {author} {\bibfnamefont
  {H.}~\bibnamefont {Ikeda}}, \ and\ \bibinfo {author} {\bibfnamefont
  {K.}~\bibnamefont {Machida}},\ }\href {\doibase
  10.1103/PhysRevLett.112.067002} {\bibfield  {journal} {\bibinfo  {journal}
  {Phys. Rev. Lett.}\ }\textbf {\bibinfo {volume} {112}},\ \bibinfo {pages}
  {067002} (\bibinfo {year} {2014})}\BibitemShut {NoStop}%
\bibitem [{\citenamefont {Shimizu}\ \emph {et~al.}(2015)\citenamefont
  {Shimizu}, \citenamefont {Kittaka}, \citenamefont {Sakakibara}, \citenamefont
  {Haga}, \citenamefont {Yamamoto}, \citenamefont {Amitsuka}, \citenamefont
  {Tsutsumi},\ and\ \citenamefont {Machida}}]{Shimizu15}%
  \BibitemOpen
  \bibfield  {author} {\bibinfo {author} {\bibfnamefont {Y.}~\bibnamefont
  {Shimizu}}, \bibinfo {author} {\bibfnamefont {S.}~\bibnamefont {Kittaka}},
  \bibinfo {author} {\bibfnamefont {T.}~\bibnamefont {Sakakibara}}, \bibinfo
  {author} {\bibfnamefont {Y.}~\bibnamefont {Haga}}, \bibinfo {author}
  {\bibfnamefont {E.}~\bibnamefont {Yamamoto}}, \bibinfo {author}
  {\bibfnamefont {H.}~\bibnamefont {Amitsuka}}, \bibinfo {author}
  {\bibfnamefont {Y.}~\bibnamefont {Tsutsumi}}, \ and\ \bibinfo {author}
  {\bibfnamefont {K.}~\bibnamefont {Machida}},\ }\href {\doibase
  10.1103/PhysRevLett.114.147002} {\bibfield  {journal} {\bibinfo  {journal}
  {Phys. Rev. Lett.}\ }\textbf {\bibinfo {volume} {114}},\ \bibinfo {pages}
  {147002} (\bibinfo {year} {2015})}\BibitemShut {NoStop}%
\bibitem [{\citenamefont {Hedo}\ \emph {et~al.}(1998)\citenamefont {Hedo},
  \citenamefont {Inada}, \citenamefont {Yamamoto}, \citenamefont {Haga},
  \citenamefont {Onuki}, \citenamefont {Aoki}, \citenamefont {D.~Matsuda},
  \citenamefont {Sato},\ and\ \citenamefont {Takahashi}}]{Hedo98}%
  \BibitemOpen
  \bibfield  {author} {\bibinfo {author} {\bibfnamefont {M.}~\bibnamefont
  {Hedo}}, \bibinfo {author} {\bibfnamefont {Y.}~\bibnamefont {Inada}},
  \bibinfo {author} {\bibfnamefont {E.}~\bibnamefont {Yamamoto}}, \bibinfo
  {author} {\bibfnamefont {Y.}~\bibnamefont {Haga}}, \bibinfo {author}
  {\bibfnamefont {Y.}~\bibnamefont {Onuki}}, \bibinfo {author} {\bibfnamefont
  {Y.}~\bibnamefont {Aoki}}, \bibinfo {author} {\bibfnamefont {T.}~\bibnamefont
  {D.~Matsuda}}, \bibinfo {author} {\bibfnamefont {H.}~\bibnamefont {Sato}}, \
  and\ \bibinfo {author} {\bibfnamefont {S.}~\bibnamefont {Takahashi}},\ }\href
  {\doibase 10.1143/JPSJ.67.272} {\bibfield  {journal} {\bibinfo  {journal} {J.
  Phys. Soc. Jpn.}\ }\textbf {\bibinfo {volume} {67}},\ \bibinfo {pages} {272}
  (\bibinfo {year} {1998})}\BibitemShut {NoStop}%
\bibitem [{\citenamefont {Mott}(1961)}]{Mott61}%
  \BibitemOpen
  \bibfield  {author} {\bibinfo {author} {\bibfnamefont {N.~F.}\ \bibnamefont
  {Mott}},\ }\href {\doibase 10.1080/14786436108243318} {\bibfield  {journal}
  {\bibinfo  {journal} {Philos. Mag.}\ }\textbf {\bibinfo {volume} {6}},\
  \bibinfo {pages} {287} (\bibinfo {year} {1961})}\BibitemShut {NoStop}%
\bibitem [{\citenamefont {Halperin}\ and\ \citenamefont
  {Rice}(1968)}]{Halperin68}%
  \BibitemOpen
  \bibfield  {author} {\bibinfo {author} {\bibfnamefont {B.~I.}\ \bibnamefont
  {Halperin}}\ and\ \bibinfo {author} {\bibfnamefont {T.~M.}\ \bibnamefont
  {Rice}},\ }\href {\doibase 10.1103/RevModPhys.40.755} {\bibfield  {journal}
  {\bibinfo  {journal} {Rev. Mod. Phys.}\ }\textbf {\bibinfo {volume} {40}},\
  \bibinfo {pages} {755} (\bibinfo {year} {1968})}\BibitemShut {NoStop}%
\bibitem [{\citenamefont {Kune$\check{\mathrm{s}}$}(2015)}]{Kunes15}%
  \BibitemOpen
  \bibfield  {author} {\bibinfo {author} {\bibfnamefont {J.}~\bibnamefont
  {Kune$\check{\mathrm{s}}$}},\ }\href
  {http://stacks.iop.org/0953-8984/27/i=33/a=333201} {\bibfield  {journal}
  {\bibinfo  {journal} {J. Phys. Condens. Matter}\ }\textbf {\bibinfo {volume}
  {27}},\ \bibinfo {pages} {333201} (\bibinfo {year} {2015})}\BibitemShut
  {NoStop}%
\bibitem [{\citenamefont {Emery}\ and\ \citenamefont
  {Kivelson}(1992)}]{Emery92}%
  \BibitemOpen
  \bibfield  {author} {\bibinfo {author} {\bibfnamefont {V.~J.}\ \bibnamefont
  {Emery}}\ and\ \bibinfo {author} {\bibfnamefont {S.}~\bibnamefont
  {Kivelson}},\ }\href {\doibase 10.1103/PhysRevB.46.10812} {\bibfield
  {journal} {\bibinfo  {journal} {Phys. Rev. B}\ }\textbf {\bibinfo {volume}
  {46}},\ \bibinfo {pages} {10812} (\bibinfo {year} {1992})}\BibitemShut
  {NoStop}%
\bibitem [{\citenamefont {Balatsky}\ and\ \citenamefont
  {Bon$\check{\mathrm{c}}$a}(1993)}]{Balatsky93}%
  \BibitemOpen
  \bibfield  {author} {\bibinfo {author} {\bibfnamefont {A.~V.}\ \bibnamefont
  {Balatsky}}\ and\ \bibinfo {author} {\bibfnamefont {J.}~\bibnamefont
  {Bon$\check{\mathrm{c}}$a}},\ }\href {\doibase 10.1103/PhysRevB.48.7445}
  {\bibfield  {journal} {\bibinfo  {journal} {Phys. Rev. B}\ }\textbf {\bibinfo
  {volume} {48}},\ \bibinfo {pages} {7445} (\bibinfo {year}
  {1993})}\BibitemShut {NoStop}%
\bibitem [{\citenamefont {Coleman}\ \emph {et~al.}(1993)\citenamefont
  {Coleman}, \citenamefont {Miranda},\ and\ \citenamefont
  {Tsvelik}}]{Coleman93}%
  \BibitemOpen
  \bibfield  {author} {\bibinfo {author} {\bibfnamefont {P.}~\bibnamefont
  {Coleman}}, \bibinfo {author} {\bibfnamefont {E.}~\bibnamefont {Miranda}}, \
  and\ \bibinfo {author} {\bibfnamefont {A.}~\bibnamefont {Tsvelik}},\ }\href
  {\doibase 10.1103/PhysRevLett.70.2960} {\bibfield  {journal} {\bibinfo
  {journal} {Phys. Rev. Lett.}\ }\textbf {\bibinfo {volume} {70}},\ \bibinfo
  {pages} {2960} (\bibinfo {year} {1993})}\BibitemShut {NoStop}%
\bibitem [{\citenamefont {Coleman}\ \emph {et~al.}(1995)\citenamefont
  {Coleman}, \citenamefont {Miranda},\ and\ \citenamefont
  {Tsvelik}}]{Coleman95}%
  \BibitemOpen
  \bibfield  {author} {\bibinfo {author} {\bibfnamefont {P.}~\bibnamefont
  {Coleman}}, \bibinfo {author} {\bibfnamefont {E.}~\bibnamefont {Miranda}}, \
  and\ \bibinfo {author} {\bibfnamefont {A.}~\bibnamefont {Tsvelik}},\ }\href
  {\doibase 10.1103/PhysRevLett.74.1653} {\bibfield  {journal} {\bibinfo
  {journal} {Phys. Rev. Lett.}\ }\textbf {\bibinfo {volume} {74}},\ \bibinfo
  {pages} {1653} (\bibinfo {year} {1995})}\BibitemShut {NoStop}%
\bibitem [{\citenamefont {Emery}\ and\ \citenamefont
  {Kivelson}(1993)}]{Emery93}%
  \BibitemOpen
  \bibfield  {author} {\bibinfo {author} {\bibfnamefont {V.~J.}\ \bibnamefont
  {Emery}}\ and\ \bibinfo {author} {\bibfnamefont {S.~A.}\ \bibnamefont
  {Kivelson}},\ }\href {\doibase 10.1103/PhysRevLett.71.3701} {\bibfield
  {journal} {\bibinfo  {journal} {Phys. Rev. Lett.}\ }\textbf {\bibinfo
  {volume} {71}},\ \bibinfo {pages} {3701} (\bibinfo {year}
  {1993})}\BibitemShut {NoStop}%
\bibitem [{\citenamefont {Coleman}\ \emph {et~al.}(1999)\citenamefont
  {Coleman}, \citenamefont {Tsvelik}, \citenamefont {Andrei},\ and\
  \citenamefont {Kee}}]{Coleman99}%
  \BibitemOpen
  \bibfield  {author} {\bibinfo {author} {\bibfnamefont {P.}~\bibnamefont
  {Coleman}}, \bibinfo {author} {\bibfnamefont {A.~M.}\ \bibnamefont
  {Tsvelik}}, \bibinfo {author} {\bibfnamefont {N.}~\bibnamefont {Andrei}}, \
  and\ \bibinfo {author} {\bibfnamefont {H.~Y.}\ \bibnamefont {Kee}},\ }\href
  {\doibase 10.1103/PhysRevB.60.3608} {\bibfield  {journal} {\bibinfo
  {journal} {Phys. Rev. B}\ }\textbf {\bibinfo {volume} {60}},\ \bibinfo
  {pages} {3608} (\bibinfo {year} {1999})}\BibitemShut {NoStop}%
\bibitem [{\citenamefont {Jarrell}\ \emph {et~al.}(1997)\citenamefont
  {Jarrell}, \citenamefont {Pang},\ and\ \citenamefont {Cox}}]{Jarrell97}%
  \BibitemOpen
  \bibfield  {author} {\bibinfo {author} {\bibfnamefont {M.}~\bibnamefont
  {Jarrell}}, \bibinfo {author} {\bibfnamefont {H.}~\bibnamefont {Pang}}, \
  and\ \bibinfo {author} {\bibfnamefont {D.~L.}\ \bibnamefont {Cox}},\ }\href
  {\doibase 10.1103/PhysRevLett.78.1996} {\bibfield  {journal} {\bibinfo
  {journal} {Phys. Rev. Lett.}\ }\textbf {\bibinfo {volume} {78}},\ \bibinfo
  {pages} {1996} (\bibinfo {year} {1997})}\BibitemShut {NoStop}%
\bibitem [{\citenamefont {Anders}(2002{\natexlab{a}})}]{Anders02}%
  \BibitemOpen
  \bibfield  {author} {\bibinfo {author} {\bibfnamefont {F.~B.}\ \bibnamefont
  {Anders}},\ }\href {\doibase 10.1103/PhysRevB.66.020504} {\bibfield
  {journal} {\bibinfo  {journal} {Phys. Rev. B}\ }\textbf {\bibinfo {volume}
  {66}},\ \bibinfo {pages} {020504} (\bibinfo {year}
  {2002}{\natexlab{a}})}\BibitemShut {NoStop}%
\bibitem [{\citenamefont {Anders}(2002{\natexlab{b}})}]{Anders02-2}%
  \BibitemOpen
  \bibfield  {author} {\bibinfo {author} {\bibfnamefont {F.~B.}\ \bibnamefont
  {Anders}},\ }\href {\doibase 10.1140/epjb/e2002-00195-8} {\bibfield
  {journal} {\bibinfo  {journal} {Eur. Phys. J. B}\ }\textbf {\bibinfo {volume}
  {28}},\ \bibinfo {pages} {9} (\bibinfo {year}
  {2002}{\natexlab{b}})}\BibitemShut {NoStop}%
\bibitem [{\citenamefont {Flint}\ \emph {et~al.}(2008)\citenamefont {Flint},
  \citenamefont {Dzero},\ and\ \citenamefont {Coleman}}]{Flint08}%
  \BibitemOpen
  \bibfield  {author} {\bibinfo {author} {\bibfnamefont {R.}~\bibnamefont
  {Flint}}, \bibinfo {author} {\bibfnamefont {M.}~\bibnamefont {Dzero}}, \ and\
  \bibinfo {author} {\bibfnamefont {P.}~\bibnamefont {Coleman}},\ }\href
  {http://dx.doi.org/10.1038/nphys1024} {\bibfield  {journal} {\bibinfo
  {journal} {Nat. Phys.}\ }\textbf {\bibinfo {volume} {4}},\ \bibinfo {pages}
  {643} (\bibinfo {year} {2008})}\BibitemShut {NoStop}%
\bibitem [{\citenamefont {Flint}\ and\ \citenamefont
  {Coleman}(2010)}]{Flint10}%
  \BibitemOpen
  \bibfield  {author} {\bibinfo {author} {\bibfnamefont {R.}~\bibnamefont
  {Flint}}\ and\ \bibinfo {author} {\bibfnamefont {P.}~\bibnamefont
  {Coleman}},\ }\href {\doibase 10.1103/PhysRevLett.105.246404} {\bibfield
  {journal} {\bibinfo  {journal} {Phys. Rev. Lett.}\ }\textbf {\bibinfo
  {volume} {105}},\ \bibinfo {pages} {246404} (\bibinfo {year}
  {2010})}\BibitemShut {NoStop}%
\bibitem [{\citenamefont {Flint}\ \emph {et~al.}(2011)\citenamefont {Flint},
  \citenamefont {Nevidomskyy},\ and\ \citenamefont {Coleman}}]{Flint14}%
  \BibitemOpen
  \bibfield  {author} {\bibinfo {author} {\bibfnamefont {R.}~\bibnamefont
  {Flint}}, \bibinfo {author} {\bibfnamefont {A.~H.}\ \bibnamefont
  {Nevidomskyy}}, \ and\ \bibinfo {author} {\bibfnamefont {P.}~\bibnamefont
  {Coleman}},\ }\href {\doibase 10.1103/PhysRevB.84.064514} {\bibfield
  {journal} {\bibinfo  {journal} {Phys. Rev. B}\ }\textbf {\bibinfo {volume}
  {84}},\ \bibinfo {pages} {064514} (\bibinfo {year} {2011})}\BibitemShut
  {NoStop}%
\bibitem [{\citenamefont {Hoshino}\ and\ \citenamefont
  {Kuramoto}(2014)}]{Hoshino14}%
  \BibitemOpen
  \bibfield  {author} {\bibinfo {author} {\bibfnamefont {S.}~\bibnamefont
  {Hoshino}}\ and\ \bibinfo {author} {\bibfnamefont {Y.}~\bibnamefont
  {Kuramoto}},\ }\href {\doibase 10.1103/PhysRevLett.112.167204} {\bibfield
  {journal} {\bibinfo  {journal} {Phys. Rev. Lett.}\ }\textbf {\bibinfo
  {volume} {112}},\ \bibinfo {pages} {167204} (\bibinfo {year}
  {2014})}\BibitemShut {NoStop}%
\bibitem [{\citenamefont {{For a recent review, see Y. Zhou, K. Kanoda, and
  T.-K. Ng}}(2017)}]{Zhou17}%
  \BibitemOpen
  \bibfield  {author} {\bibinfo {author} {\bibnamefont {{For a recent review,
  see Y. Zhou, K. Kanoda, and T.-K. Ng}}},\ }\href {\doibase
  10.1103/RevModPhys.89.025003} {\bibfield  {journal} {\bibinfo  {journal}
  {Rev. Mod. Phys.}\ }\textbf {\bibinfo {volume} {89}},\ \bibinfo {pages}
  {025003} (\bibinfo {year} {2017})}\BibitemShut {NoStop}%
\bibitem [{\citenamefont {{For a recent review, see L. Savary and L.
  Balents}}(2017)}]{Savary17}%
  \BibitemOpen
  \bibfield  {author} {\bibinfo {author} {\bibnamefont {{For a recent review,
  see L. Savary and L. Balents}}},\ }\href
  {http://stacks.iop.org/0034-4885/80/i=1/a=016502} {\bibfield  {journal}
  {\bibinfo  {journal} {Rep. Prog. Phys.}\ }\textbf {\bibinfo {volume} {80}},\
  \bibinfo {pages} {016502} (\bibinfo {year} {2017})}\BibitemShut {NoStop}%
\bibitem [{\citenamefont {Cox}(1987)}]{Cox87}%
  \BibitemOpen
  \bibfield  {author} {\bibinfo {author} {\bibfnamefont {D.~L.}\ \bibnamefont
  {Cox}},\ }\href {\doibase 10.1103/PhysRevLett.59.1240} {\bibfield  {journal}
  {\bibinfo  {journal} {Phys. Rev. Lett.}\ }\textbf {\bibinfo {volume} {59}},\
  \bibinfo {pages} {1240} (\bibinfo {year} {1987})}\BibitemShut {NoStop}%
\bibitem [{\citenamefont {Cox}\ and\ \citenamefont {Zawadowski}(1998)}]{Cox98}%
  \BibitemOpen
  \bibfield  {author} {\bibinfo {author} {\bibfnamefont {D.~L.}\ \bibnamefont
  {Cox}}\ and\ \bibinfo {author} {\bibfnamefont {A.}~\bibnamefont
  {Zawadowski}},\ }\href {\doibase 10.1080/000187398243500} {\bibfield
  {journal} {\bibinfo  {journal} {Adv. Phys.}\ }\textbf {\bibinfo {volume}
  {47}},\ \bibinfo {pages} {599} (\bibinfo {year} {1998})}\BibitemShut
  {NoStop}%
\bibitem [{\citenamefont {Hoshino}(2014)}]{Hoshino14-2}%
  \BibitemOpen
  \bibfield  {author} {\bibinfo {author} {\bibfnamefont {S.}~\bibnamefont
  {Hoshino}},\ }\href {\doibase 10.1103/PhysRevB.90.115154} {\bibfield
  {journal} {\bibinfo  {journal} {Phys. Rev. B}\ }\textbf {\bibinfo {volume}
  {90}},\ \bibinfo {pages} {115154} (\bibinfo {year} {2014})}\BibitemShut
  {NoStop}%
\bibitem [{\citenamefont {Kusunose}(2016)}]{Kusunose16}%
  \BibitemOpen
  \bibfield  {author} {\bibinfo {author} {\bibfnamefont {H.}~\bibnamefont
  {Kusunose}},\ }\href {\doibase 10.7566/JPSJ.85.113701} {\bibfield  {journal}
  {\bibinfo  {journal} {J. Phys. Soc. Jpn.}\ }\textbf {\bibinfo {volume}
  {85}},\ \bibinfo {pages} {113701} (\bibinfo {year} {2016})}\BibitemShut
  {NoStop}%
\bibitem [{\citenamefont {Zhang}\ and\ \citenamefont {Yu}(2000)}]{Zhang00}%
  \BibitemOpen
  \bibfield  {author} {\bibinfo {author} {\bibfnamefont {G.-M.}\ \bibnamefont
  {Zhang}}\ and\ \bibinfo {author} {\bibfnamefont {L.}~\bibnamefont {Yu}},\
  }\href {\doibase 10.1103/PhysRevB.62.76} {\bibfield  {journal} {\bibinfo
  {journal} {Phys. Rev. B}\ }\textbf {\bibinfo {volume} {62}},\ \bibinfo
  {pages} {76} (\bibinfo {year} {2000})}\BibitemShut {NoStop}%
\bibitem [{\citenamefont {Capponi}\ and\ \citenamefont
  {Assaad}(2001)}]{Capponi01}%
  \BibitemOpen
  \bibfield  {author} {\bibinfo {author} {\bibfnamefont {S.}~\bibnamefont
  {Capponi}}\ and\ \bibinfo {author} {\bibfnamefont {F.~F.}\ \bibnamefont
  {Assaad}},\ }\href {\doibase 10.1103/PhysRevB.63.155114} {\bibfield
  {journal} {\bibinfo  {journal} {Phys. Rev. B}\ }\textbf {\bibinfo {volume}
  {63}},\ \bibinfo {pages} {155114} (\bibinfo {year} {2001})}\BibitemShut
  {NoStop}%
\bibitem [{\citenamefont {Kubo}(1957)}]{Kubo57}%
  \BibitemOpen
  \bibfield  {author} {\bibinfo {author} {\bibfnamefont {R.}~\bibnamefont
  {Kubo}},\ }\href {\doibase 10.1143/JPSJ.12.570} {\bibfield  {journal}
  {\bibinfo  {journal} {J. Phys. Soc. Jpn.}\ }\textbf {\bibinfo {volume}
  {12}},\ \bibinfo {pages} {570} (\bibinfo {year} {1957})}\BibitemShut
  {NoStop}%
\bibitem [{\citenamefont {Georges}\ \emph {et~al.}(1996)\citenamefont
  {Georges}, \citenamefont {Kotliar}, \citenamefont {Krauth},\ and\
  \citenamefont {Rozenberg}}]{Georges96}%
  \BibitemOpen
  \bibfield  {author} {\bibinfo {author} {\bibfnamefont {A.}~\bibnamefont
  {Georges}}, \bibinfo {author} {\bibfnamefont {G.}~\bibnamefont {Kotliar}},
  \bibinfo {author} {\bibfnamefont {W.}~\bibnamefont {Krauth}}, \ and\ \bibinfo
  {author} {\bibfnamefont {M.~J.}\ \bibnamefont {Rozenberg}},\ }\href {\doibase
  10.1103/RevModPhys.68.13} {\bibfield  {journal} {\bibinfo  {journal} {Rev.
  Mod. Phys.}\ }\textbf {\bibinfo {volume} {68}},\ \bibinfo {pages} {13}
  (\bibinfo {year} {1996})}\BibitemShut {NoStop}%
\bibitem [{\citenamefont {Khurana}(1990)}]{Khurana90}%
  \BibitemOpen
  \bibfield  {author} {\bibinfo {author} {\bibfnamefont {A.}~\bibnamefont
  {Khurana}},\ }\href {\doibase 10.1103/PhysRevLett.64.1990} {\bibfield
  {journal} {\bibinfo  {journal} {Phys. Rev. Lett.}\ }\textbf {\bibinfo
  {volume} {64}},\ \bibinfo {pages} {1990} (\bibinfo {year}
  {1990})}\BibitemShut {NoStop}%
\bibitem [{\citenamefont {Mattheiss}(1973)}]{Mattheiss73}%
  \BibitemOpen
  \bibfield  {author} {\bibinfo {author} {\bibfnamefont {L.~F.}\ \bibnamefont
  {Mattheiss}},\ }\href {\doibase 10.1103/PhysRevB.8.3719} {\bibfield
  {journal} {\bibinfo  {journal} {Phys. Rev. B}\ }\textbf {\bibinfo {volume}
  {8}},\ \bibinfo {pages} {3719} (\bibinfo {year} {1973})}\BibitemShut
  {NoStop}%
\bibitem [{\citenamefont {Onimaru}\ \emph {et~al.}(2011)\citenamefont
  {Onimaru}, \citenamefont {Matsumoto}, \citenamefont {Inoue}, \citenamefont
  {Umeo}, \citenamefont {Sakakibara}, \citenamefont {Karaki}, \citenamefont
  {Kubota},\ and\ \citenamefont {Takabatake}}]{Onimaru11}%
  \BibitemOpen
  \bibfield  {author} {\bibinfo {author} {\bibfnamefont {T.}~\bibnamefont
  {Onimaru}}, \bibinfo {author} {\bibfnamefont {K.~T.}\ \bibnamefont
  {Matsumoto}}, \bibinfo {author} {\bibfnamefont {Y.~F.}\ \bibnamefont
  {Inoue}}, \bibinfo {author} {\bibfnamefont {K.}~\bibnamefont {Umeo}},
  \bibinfo {author} {\bibfnamefont {T.}~\bibnamefont {Sakakibara}}, \bibinfo
  {author} {\bibfnamefont {Y.}~\bibnamefont {Karaki}}, \bibinfo {author}
  {\bibfnamefont {M.}~\bibnamefont {Kubota}}, \ and\ \bibinfo {author}
  {\bibfnamefont {T.}~\bibnamefont {Takabatake}},\ }\href {\doibase
  10.1103/PhysRevLett.106.177001} {\bibfield  {journal} {\bibinfo  {journal}
  {Phys. Rev. Lett.}\ }\textbf {\bibinfo {volume} {106}},\ \bibinfo {pages}
  {177001} (\bibinfo {year} {2011})}\BibitemShut {NoStop}%
\bibitem [{\citenamefont {Loucks}\ and\ \citenamefont
  {Cutler}(1964)}]{Loucks64}%
  \BibitemOpen
  \bibfield  {author} {\bibinfo {author} {\bibfnamefont {T.~L.}\ \bibnamefont
  {Loucks}}\ and\ \bibinfo {author} {\bibfnamefont {P.~H.}\ \bibnamefont
  {Cutler}},\ }\href {\doibase 10.1103/PhysRev.133.A819} {\bibfield  {journal}
  {\bibinfo  {journal} {Phys. Rev.}\ }\textbf {\bibinfo {volume} {133}},\
  \bibinfo {pages} {A819} (\bibinfo {year} {1964})}\BibitemShut {NoStop}%
\bibitem [{\citenamefont {Inoue}\ and\ \citenamefont
  {Yamashita}(1973)}]{Inoue73}%
  \BibitemOpen
  \bibfield  {author} {\bibinfo {author} {\bibfnamefont {S.~T.}\ \bibnamefont
  {Inoue}}\ and\ \bibinfo {author} {\bibfnamefont {J.}~\bibnamefont
  {Yamashita}},\ }\href {\doibase 10.1143/JPSJ.35.677} {\bibfield  {journal}
  {\bibinfo  {journal} {J. Phys. Soc. Jpn.}\ }\textbf {\bibinfo {volume}
  {35}},\ \bibinfo {pages} {677} (\bibinfo {year} {1973})}\BibitemShut
  {NoStop}%
\bibitem [{\citenamefont {Yamane}\ \emph {et~al.}(2018)\citenamefont {Yamane},
  \citenamefont {Onimaru}, \citenamefont {Wakiya}, \citenamefont {Matsumoto},
  \citenamefont {Umeo},\ and\ \citenamefont {Takabatake}}]{Yamane18}%
  \BibitemOpen
  \bibfield  {author} {\bibinfo {author} {\bibfnamefont {Y.}~\bibnamefont
  {Yamane}}, \bibinfo {author} {\bibfnamefont {T.}~\bibnamefont {Onimaru}},
  \bibinfo {author} {\bibfnamefont {K.}~\bibnamefont {Wakiya}}, \bibinfo
  {author} {\bibfnamefont {K.~T.}\ \bibnamefont {Matsumoto}}, \bibinfo {author}
  {\bibfnamefont {K.}~\bibnamefont {Umeo}}, \ and\ \bibinfo {author}
  {\bibfnamefont {T.}~\bibnamefont {Takabatake}},\ }\href {\doibase
  10.1103/PhysRevLett.121.077206} {\bibfield  {journal} {\bibinfo  {journal}
  {Phys. Rev. Lett.}\ }\textbf {\bibinfo {volume} {121}},\ \bibinfo {pages}
  {077206} (\bibinfo {year} {2018})}\BibitemShut {NoStop}%
\bibitem [{\citenamefont {Stark}\ and\ \citenamefont
  {Falicov}(1967)}]{Stark67}%
  \BibitemOpen
  \bibfield  {author} {\bibinfo {author} {\bibfnamefont {R.~W.}\ \bibnamefont
  {Stark}}\ and\ \bibinfo {author} {\bibfnamefont {L.~M.}\ \bibnamefont
  {Falicov}},\ }\href {\doibase 10.1103/PhysRevLett.19.795} {\bibfield
  {journal} {\bibinfo  {journal} {Phys. Rev. Lett.}\ }\textbf {\bibinfo
  {volume} {19}},\ \bibinfo {pages} {795} (\bibinfo {year} {1967})}\BibitemShut
  {NoStop}%
\bibitem [{\citenamefont {Allen}\ \emph {et~al.}(1968)\citenamefont {Allen},
  \citenamefont {Cohen}, \citenamefont {Falicov},\ and\ \citenamefont
  {Kasowski}}]{Allen68}%
  \BibitemOpen
  \bibfield  {author} {\bibinfo {author} {\bibfnamefont {P.~B.}\ \bibnamefont
  {Allen}}, \bibinfo {author} {\bibfnamefont {M.~L.}\ \bibnamefont {Cohen}},
  \bibinfo {author} {\bibfnamefont {L.~M.}\ \bibnamefont {Falicov}}, \ and\
  \bibinfo {author} {\bibfnamefont {R.~V.}\ \bibnamefont {Kasowski}},\ }\href
  {\doibase 10.1103/PhysRevLett.21.1794} {\bibfield  {journal} {\bibinfo
  {journal} {Phys. Rev. Lett.}\ }\textbf {\bibinfo {volume} {21}},\ \bibinfo
  {pages} {1794} (\bibinfo {year} {1968})}\BibitemShut {NoStop}%
\end{thebibliography}

\end{document}


\title{
Supplementary Materials for "Unconventional Full-Gap Superconductivity in Kondo Lattice with Semi-Metallic Conduction Bands"\textit{}
}

\author{Shoma Iimura$^{1}$} 
\author{Motoaki Hirayama$^2$}
\author{Shintaro Hoshino$^{1}$}

\affiliation{
$^1$Department of Physics, Saitama University, Shimo-Okubo, Saitama 338-8570, Japan\\
$^2$RIKEN Center for Emergent Matter Science (CEMS), Wako, Saitama 351-0198, Japan
}
\maketitle


Hereafter, we take $\hbar$ as unity.
\section{Mean-Field Approximation}
\label{MFA}
In this section, we employ the mean-field approximation and decouple the interaction between conduction electron and Kramers / non-Kramers doublet.
\subsection{Kramers Doublet System}
For the Kramers doublet, the interaction term is described by the pseudofermion operator $f_\sigma (\bm{r})$ as follows.
\begin{align}
\mathcal{H}_{\mathrm{int}} &= \frac{1}{4}\sum_{\sg_1 \sg_2}\sum_{\al,\sigma \sigma'} J_\alpha \int d{\bm r}~f_{\sg_1}^\dagger ({\bm r}){\bm \sg}_{\sg_1 \sg_2} f_{\sg_2}^{ } ({\bm r})\cdot \psi_{\alpha\sigma}^\dagger ({\bm r}) {\bm{\sigma}}_{\sigma \sigma'} \psi_{\alpha \sigma'}^{ } ({\bm r}).
\end{align}
The interaction between the electron Fermi surface and the localized spin is decoupled as
\begin{align}
\label{mean_field_hamiltonian}
\mathcal{H}_{\mathrm{int},\al=1} \simeq \sum_{\bm{k},\sigma} V^* \psi_{\bm{k}1\sigma}^\dagger f_{\bm{k}\sigma}^{ }+ \mathrm{H.c.},
\end{align}
where $V$ represents the mean-field defined in Eq.~(3). Similarly, we decouple the interaction between the hole surface and the localized spin as,
\begin{align}
\mathcal{H}_{\mathrm{int},\al=2}\simeq \sum_{\bm{k},\sigma} W^* \epsilon_{\sigma \bar{\sigma}} \psi_{-\bm{k}2 \bar{\sigma}}^{ } f_{\bm{k}\sigma}^{ } + \mathrm{H.c.},
\end{align}
where $W$ is defined in Eq.~(4).

The mean-field Hamiltonian for the Kramers doublet is given by 
\begin{align}
\label{MF_Kramers}
\mathcal{H}^{\mathrm{MF}} = \sum_{\bm{k},\sigma} \Psi_{\bm{k}\sigma}^\dagger 
\left( \begin{array}{ccc}
\xi_{\bm{k}1} & 0 & V^* \\
0 & -\xi_{\bm{k}2} & W^* \epsilon_{\sigma \bar{\sigma}} \\
V & W \epsilon_{\sigma \bar{\sigma}} & 0
\end{array} \right) 
\Psi_{ {\bm k} \sigma}^{ },
\end{align}
where the Nambu representation of the field operators is defined as follows.
\begin{align}
\Psi_{\bm{k}\sigma}^\dagger = \left( \psi_{\bm{k}1\sigma }^\dagger,\psi_{-\bm{k} 2 \bar{\sigma} }^{ },f_{\bm{k} \sigma }^\dagger \right).
\end{align}

\subsection{Non-Kramers Doublet System}
For the non-Kramers doublet, we introduce the pseudofermion operator $d_\alpha ({\bm r})$, which describes the pseudospin degrees of freedom. The localized pseudospin operator ${\bm T}({\bm r})$ can be rewritten in terms of the pseudofermion as ${\bm T}({\bm r}) = \frac{1}{2} \sum_{\alpha \alpha'} d_\alpha^\dagger ({\bm r}) {\bm \sigma}_{\al \al'} d_{\al'}^{ } ({\bm r})$.
Then, the interaction between the non-Kramers doublet and the semi-metallic conduction bands is described as follows.
\begin{align}
\mathcal H_{\rm int} &= 
\frac{1}{4} \sum_{\al_1 \al_2} \sum_{\al\al'\sigma} J \int d{\bm r}~d_{\al_1}^\dagger ({\bm r}) {\bm \sigma}_{\al_1 \al_2} d_{\al_2}^{ } ({\bm r}) \cdot \psi_{\alpha\sigma}^\dagger ({\bm r}){\bm{\sigma}}_{\al\al'} \psi_{\al' \sigma}^{ } ({\bm r}).
\end{align}
We introduce the following mean-fields,
\begin{align}
&\tilde{V} = \frac{3}{4} J \langle d_\al^{ } ({\bm r}) \psi_{\al \uparrow}^\dagger ({\bm r})\rangle, \\
&\tilde{W}\epsilon_{\al \al'} = \frac{3}{4} J \langle d_\al^{ } ({\bm r}) \psi_{\al' \downarrow}^{ } ({\bm r}) \rangle.
\end{align} 
The resultant mean-field Hamiltonian is equivalent to the Kramers case.

\subsection{Validity of mean-field theory}
We comment on the validity of our mean-field approach to the Kondo lattice with semi-metallic conduction bands.
For a typical heavy-electron system, the localized moments of $f$ electrons at high temperatures crossover to the heavy-electron state at low temperatures by a collective Kondo effect.
It is recognized that the mean-field approximation, which we have used in our paper, changes the crossover into the transition, which is an artifact of the approximation.
On the other hand, the ground state with heavy electrons is correctly reproduced by the mean-field theory, and the transition temperature predicts the crossover energy scale obtained in the more accurate method.
For example, the comparison has been done in the study of two-dimensional Kondo lattice \cite{Capponi01}. 
\\
\indent
The justification for the use of the mean-field theory has been demonstrated also in the two-channel Kondo lattice models. The dynamical mean-field theory, which takes full account of local electronic correlation necessary for the Kondo effects, has revealed several nontrivial ordered phases including unconventional superconductivity \cite{Hoshino11,Hoshino14}. These results are reasonably reproduced by the mean-field theory as demonstrated in Ref.~\cite{Hoshino14-2}. In the present work, we apply this established method to the Kondo lattice problem with semi-metallic conduction bands.

\section{Composite Order Parameter and Correlation}
\label{COP}
In this section, we evaluate the composite pair amplitude, which is defined in Eq.~(6). It can be rewritten in the mean-field theory as follows,
\begin{align}
\Phi_{\mu,\sigma\sigma'}^{\alpha \alpha'} (\bm{R};\bm{r},\bm{r}') &\simeq \frac{1}{2} \sum_{\sigma_1,\sigma_2} \sigma^\mu_{\sigma_1 \sigma_2}  \bigl[ \delta_{\alpha 1}\delta_{\alpha' 2} \delta_{\sigma_1 \sigma} F_{1\sigma}(\bm{r}-\bm{R}) F_{2 \sigma_2 \sigma'} (\bm{r}'-\bm{R})\bigr. \nonumber \\
&\hspace{7em}\bigl.  -\delta_{\alpha 2} \delta_{\alpha' 1} \delta_{\sigma_1 \sigma'} F_{1\sigma'}(\bm{r}'-\bm{R}) F_{2\sigma_2 \sigma} (\bm{r} - \bm{R}) \bigr],
\end{align}
where $F_{1\sigma}(\bm{r})$ and $F_{2\sigma \sigma'}(\bm{r})$ are defined respectively as,
\begin{align}
&F_{1\sigma} (\bm{r}) \equiv \frac{1}{\Omega} \sum_{\bm{k}} \langle \psi_{\bm{k}1\sigma}^{ } f_{\bm{k}\sigma}^\dagger \rangle \mathrm{e}^{i\bm{k} \cdot \bm{r}},\\
&F_{2\sigma \sigma'} (\bm{r}) \equiv \frac{1}{\Omega} \sum_{\bm{k}} \langle f_{\bm{k}\sigma}^{ }\psi_{-\bm{k}2\sigma'}^{ }  \rangle \mathrm{e}^{-i\bm{k} \cdot \bm{r}}.
\end{align}
In the low-temperature limit, $F_{1\sigma}(\bm{r})$ and $F_{2\sg \sg'}({\bm r})$ can be rewritten as follows,
\begin{align}
\label{eq:correlation_3}
F_{1\sigma}( \bm{r} ) &= \frac{V^*}{\Omega} \sum_{\bm{k}\in\mathcal{K}_1} \frac{f(E_{\bm{k}1-}) - f(E_{\bm{k}1+})}{E_{\bm{k}1+} - E_{\bm{k}1-}}  \mathrm{e}^{i\bm{k}\cdot \bm{r}} 
\nonumber \\
&
=\frac{V^*}{\left( 2\pi \right)^3} \int_{\bm{k}\in \mathcal{K}_1} d\bm{k} \frac{\mathrm{e}^{i\bm{k}\cdot \bm{r}}}{\sqrt{\xi_{\bm{k}1}^2 + 4\left| V \right|^2}} \equiv V^* F_{1}({\bm r}), \\
\label{eq:correlation_4}
F_{2\sigma \sigma'}( \bm{r} ) &= \frac{W\epsilon_{\sigma \sigma'}}{\Omega} \sum_{\bm{k}\in\mathcal{K}_2} \frac{f(E_{\bm{k}2-}) - f(E_{\bm{k}2+})}{E_{\bm{k}2+} - E_{\bm{k}2-}}  \mathrm{e}^{-i\bm{k}\cdot \bm{r}} 
\nonumber \\
&
=\frac{W \epsilon_{\sg\sg'}}{\left( 2\pi \right)^3} \int_{\bm{k}\in \mathcal{K}_2} d\bm{k} \frac{\mathrm{e}^{-i\bm{k}\cdot \bm{r}}}{\sqrt{\xi_{\bm{k}2}^2 + 4\left| W \right|^2}} \equiv W \epsilon_{\sg \sg'} F_{2}^{*}({\bm r}).
\end{align}
Therefore, we obtain the composite pair amplitude given in Eq.~(6).

We derive the behavior of $F_\al ({\bm r})$ given in Eqs.~(7) and (8) below. First, we integrate the angular valuables since the energy dispersion is isotropic around $\bm{K}_\al$. Then, $F_{\al}({\bm r})$ is rewritten as follows.
\begin{align}
F_{\al}(\bm{r}) &= \frac{\mathrm{e}^{i {\bm K}_\al \cdot {\bm r}} }{k_{\mathrm{F}\al} r} s_\al \int_{-\mu_\al}^{s_\al D_\al} d\epsilon \rho_\al (\epsilon) \frac{\mathrm{sin}\left( k_{\mathrm{F}\al} r \sqrt{1+ \frac{\epsilon}{\mu_\al} } \right)}{\sqrt{(1 + \frac{\epsilon}{\mu_\al})(\epsilon^2 + 4\left| V_\al \right|^2 )}},
\end{align}
where $s_\al = \sigma^z_{\al \al}$. 
$\rho_\al (\epsilon) = \sum_{\bm k} \delta(\epsilon - \xi_{{\bm k}\al} ) /\Omega$ represents the density of states (DOS) of the bare conduction electrons. 
Hereafter, we use the constant DOS for $\rho_\al (\epsilon)$ to simplify the following calculations.

At $\bm{r}=\bm{0}$, $F_{\al}(\bm{0})$ is rewritten as,
\begin{align}
F_{\al}(\bm{0}) &= s_\al \int_{-\mu_\al}^{s_\al D_\al} d\epsilon \rho_\al (\epsilon) \frac{1}{\sqrt{ \epsilon^2 + 4\left| V_\al \right|^2}} \nonumber \\
&\simeq  \rho_\al (0) \mathrm{log}\left( \frac{D_\al |\mu_\al|}{\left| V_\al \right|^2} \right).
\end{align}
On the other hand, we can obtain the following representation for $|{\bm r}|\to \infty$.
\begin{align}
F_{\al}(|\bm{r}|\to \infty) &\simeq 2\mathrm{e}^{i {\bm K}_\al \cdot {\bm r}} \rho_\al (0) \frac{\mathrm{sin}(k_{\mathrm{F}\al} r)}{k_{\mathrm{F}\al} r} K_0 \left(\frac{r}{\xi_\al}\right), 
\end{align}
where we have used $| V_\al | \ll D_\al,|\mu_\al|$. 
$K_0(z)$ is a zeroth-modified Bessel function of the second kind, and is given by 
$K_0 (z) = \int_{0}^{\infty} dx~\mathrm{cos}(zx)/\sqrt{1+x^2}$. 
Since $K_0 (z)$ behaves as $K_0(z) \sim \mathrm{e}^{-z}$ for $| z | \gg 1$, $F_\al ({\bm r})$ for $r \gg \xi_\al$ is given as Eq. (8). 

\section{Self-Consistent Equations for Mean-Fields}
\label{SCEMF}
In this section, we derive the self-consistent equation for the effective mean-fields $V$ and $W$. 
Then, the mean-field equations in Eqs.~(3) and (4) can be rewritten by utilizing the calculations in Eqs.~\eqref{eq:correlation_3} and \eqref{eq:correlation_4}, as
\begin{align}
1 = \frac{3}{4} \frac{J_\al}{\Omega} \sum_{\bm k} \frac{ f(E_{{\bm k}\al -}) - f(E_{{\bm k}\al+}) }{E_{{\bm k}\al+} - E_{{\bm k}\al -}}.
\end{align}
Now, we derive the critical temperature $T_{c,\al}$ for the mean-field $V_\al$. 

At the critical temperature, the amplitude of the mean-field becomes 0, and the self-consistent equation is rewritten as follows,
\begin{align}
1 \simeq \frac{3}{8} J_\al \rho_\al (0) \int_{-|\mu_\al|}^{D_\al} d\epsilon \frac{\mathrm{tanh}\left( \frac{\left| \epsilon \right|}{2 k_{\mathrm{B}} T_c} \right)}{\left| \epsilon \right|}.
\end{align}
The integration is similar to that appears in the BCS theory. Thus we can obtain the critical temperature in Eq.~(10).

On the other hand, since the self-consistent equation is rewritten as $1 \simeq (3 J_\al /4) F_\al ({\bm 0})$ at zero temperature, we obtain the amplitude of the mean-field as,
\begin{align}
\left| V_\al (T=0) \right| \simeq \sqrt{D_\al \left| \mu_\al \right|}~\mathrm{exp}\left[ - \frac{2}{3 J_\al \rho_\al (0)} \right].
\end{align}

From above, the ratio between $T_{c,\al}$ and the amplitude of the energy gap $E_{\mathrm{gap},\al} = |V_\al|^2/|\mu_\al| + |V_\al|^2/D_\al$ at zero temperature is given by,
\begin{align}
\frac{E_{\mathrm{gap},\al}}{k_{\mathrm{B}} T_{c,\al}} = \frac{\pi}{2 \mathrm{e}^\gamma} \times \frac{D_\al+|\mu_\al|}{\sqrt{D_\al |\mu_\al|}}.
\end{align}
In the above calculations, we have used the constant DOS for simplicity. Once the energy dependence of the DOS is considered, the constant in Eq.~(10) is modified from 1.13. This is due to the fact that the Fermi energy $\mu_\al$ and energy range $D_\al$ have same orders of magnitude. In contrast, for BCS superconductors, the energy scales of Fermi energy and Debye frequency are much separated, so the prefactor in the expression of $T_c$ is universal.

\section{Induced Pair Amplitude}
\label{IPA}
Now, we consider the band cut-off at the high-energy $E_c$, which is regarded as a lattice-regularization. If $E_c$ is finite, the anomalous part of the electron Green's function ${\tilde{\mathcal{F}}}_{{\bm k},\sg\sg'}^{\al \al'}(i\omega_n) = \int_0^\beta d\tau \la -T_\tau \bigl[ \psi_{{\bm k}\al \sg} (\tau)\psi_{-{\bm k}\al'\sg'} \bigr] \ra \mathrm{e}^{i\omega_n \tau}$, where $\omega_n = (2 n+1)\pi/\beta$ is fermionic Matsubara frequency is given as follows,
\begin{align}
\label{eq:induced_pair}
{\tilde{\mathcal{F}}}_{\bm{k},\sigma \sigma'}^{\alpha \alpha'} (i\omega_n) &= \epsilon_{\al \al'} \epsilon_{\sg \sg'} \frac{s_\al V^* W}{\bar{D}(s_\al i\omega_n)},
\end{align}
where $\bar{D}(z)$ is a function of a complex number $z$. It is defined as $\bar{D}(z)= z y_1 y_2 - |V|^2 y_2 - |W|^2 y_1$, where $y_\al = z - s_\al \xi_{{\bm k}\al}$. When ${\bm k} \in \mathcal{K}_\al$, $|\xi_{{\bm k}\bar{\al}}|$ is replaced with $E_c$ ($\gg |\xi_{{\bm k}\al}|,|V|,|W|$), and ${\tilde{\mathcal{F}}}_{{\bm k},\sg \sg'}^{\al \al'} (i \omega_n)$ is rewritten as follows,
\begin{align}
{\tilde{\mathcal{F}}}_{\bm{k},\sigma \sigma'}^{\alpha \alpha'} (i\omega_n) &\simeq  -\epsilon_{\sigma\sigma'} \frac{V^*W}{E_c} \left[ \sg^x_{\alpha\alpha'} F_{\bm{k}}^e (i\omega_n ) + \epsilon_{\alpha\alpha'} F_{\bm{k}}^o (i \omega_n) \right],
\end{align}
with
\begin{align}
\label{eq:even_pair}
F_{\bm{k}}^{e} (i\omega_n) &= \frac{1}{2} \sum_{\Lambda=\pm 1} \sum_\alpha \frac{\theta\left( k_c - \left| \bm{k}-\bm{K}_\alpha\right| \right)}{\left(i\omega_n -\Lambda s_\al E_{\bm{k}\alpha+} \right)\left( i\omega_n -\Lambda s_\al E_{\bm{k}\alpha-} \right)}, \\
\label{eq:odd_pair}
F_{\bm{k}}^{o} (i\omega_n ) &= \frac{1}{2} \sum_{\Lambda=\pm 1} \sum_\alpha \frac{\Lambda \theta\left( k_c - \left| \bm{k}-\bm{K}_\alpha\right| \right)}{\left(i\omega_n -\Lambda s_\al E_{\bm{k}\alpha+} \right)\left( i\omega_n -\Lambda s_\al E_{\bm{k}\alpha-} \right)},
\end{align}
where $F_{\bm{k}}^e (i\omega_n)$ and $F_{\bm{k}}^o (i\omega_n)$ are even and odd with respect to the Matsubara frequency, respectively.

\section{Meissner Response}
\label{MR}

In this section, we consider the electromagnetic response in the superconducting state. Generally, we can rewrite the electromagnetic kernel in terms of the Green's function as follows, 
\begin{align}
K_{\mu\nu}^p (\bm{q},\nu_m) &= e^2 \int \frac{d \bm{k}}{\left( 2\pi\right)^3}~\frac{1}{\beta}\sum_{n=-\infty}^\infty \sum_{\alpha \alpha'}\sum_{\sigma \sigma'}
\left[v_{\bm{k}\alpha}^\mu v_{\bm{k}\alpha'}^\nu \tilde{\mathcal{G}^{ }}_{\bm{k}_+,\sigma\sigma'}^{\alpha\alpha'} (i\omega_n+i\nu_m) \tilde{\mathcal{G}^{ }}_{\bm{k}_-,\sigma \sigma'}^{\alpha \alpha'} (i\omega_n) \right.\nonumber \\
&\hspace{14em}
\left. - v_{\bm{k}\alpha}^\mu v_{-\bm{k}\alpha'}^\nu {\tilde{\mathcal{F}^{ }}}_{\bm{k}_+,\sigma \sigma'}^{\alpha \alpha'} (i\omega_{n}+i\nu_m) {\tilde{\mathcal{F}^\dagger}}_{\bm{k}_-,\sigma \sigma'}^{\alpha \alpha'} (i\omega_n)\right],\\
K_{\mu\nu}^d (\bm{q},\nu_m) &=\delta_{\mu\nu}\delta_{m0} e^2 \int \frac{d{\bm{k}}}{\left( 2\pi \right)^3}  \sum_{\alpha,\sigma} \frac{\partial v_{\bm{k}\alpha}^\mu}{\partial k_\mu} \frac{1}{\beta} \sum_{n=-\infty}^\infty \tilde{\mathcal{G}^{ }}_{\bm{k},\sigma \sigma}^{\alpha \alpha} (i\omega_n),
\end{align}
where $v_{\bm{k}\alpha}^\mu = \partial \varepsilon_{\bm{k}\alpha}/\partial k_\mu$ denotes the velocity operator. $\nu_m = 2\pi m/\beta$ is bosonic Matsubara frequency. $\tilde{\mathcal{G}}(z)$ represents the electron Green's function including influences of all orders of the lattice-regularization. It is defined as follows
\begin{align}
\tilde{\mathcal{G}}_{\bm{k}}^{-1} (z)= \mathcal{G}_{\bm{k}}^{-1} (z)- {\tilde{\Sigma}}_{\bm{k}} (z),
\end{align}
where $\mathcal{G}_{\bm{k},\sg \sg'}^{\al \al'} (i\omega_n) = \int_0^\beta d\tau \bigl< -T_\tau \bigl[ \psi_{\bm{k}\al\sg}^{ }(\tau) \psi_{\bm{k}\al'\sg'}^\dagger (0)\bigr]\bigr>\mathrm{e}^{i\omega_n \tau}$ describes the electron Green's function in the low-energy effective theory. $\tilde{\Sigma}_{\bm{k}} (z)$ denotes a self-energy, which describes the contribution from the lattice-regularization and is given as,
\begin{align}
{\tilde{\Sigma^{ }}}_{\bm{k},\sigma \sigma'}^{\alpha \alpha'} (z) = \delta_{\alpha\alpha'} \delta_{\sigma \sigma'}\frac{\left| V \right|^2 \left| W \right|^2 }{z^2} {\mathcal{G}^{ }}_{\bm{k},\sigma \sigma}^{\bar{\alpha} \bar{\alpha}} (z).
\end{align}

We derive the relation on the partial derivative of the Green's function $\tilde{\mathcal{G}}$ by the analytical calculation. 
It is given as follows,
\begin{align}
\frac{\partial}{\partial k_\mu} \tilde{\mathcal{G}^{ }}_{\bm{k},\sigma \sigma}^{\alpha \alpha'} (z) &= \left( \tilde{\mathcal{G}^{ }}_{\bm{k},\sigma \sigma}^{\alpha \alpha} (z)\right)^2 v_{\bm{k}\alpha}^\mu 
+ {\tilde{\mathcal{F}^{ }}}_{\bm{k},\sigma \bar{\sigma}}^{\alpha \bar{\alpha}} (z) {\tilde{{\mathcal{F}^\dagger}}}_{\bm{k},\sigma \bar{\sigma}}^{\alpha \bar{\alpha}} (z) v_{-\bm{k}\bar{\alpha}}^\mu.
\end{align}
Therefore, the diamagnetic term in the static limit $\bm{q}\to\bm{0}$ and $\nu_m=0$ can be transformed by calculating the partial integration. 
\begin{align}
K_{\mu\nu}^d (\bm{q},0) &= - \delta_{\mu\nu} e^2 \int \frac{d\bm{k}}{(2\pi)^3}~\frac{1}{\beta} \sum_{m=-\infty}^\infty \sum_{\alpha,\sigma} 
\left[ v_{\bm{k}\alpha}^\mu v_{\bm{k}\alpha}^\mu \left( \tilde{\mathcal{G}^{ }}_{\bm{k},\sigma \sigma}^{\alpha \alpha} (i\omega_m) \right)^2 \right. 
\left.+ v_{\bm{k}\alpha}^\mu v_{-\bm{k}\bar{\alpha}}^\mu {\tilde{\mathcal{F}^{ }}}_{\bm{k},\sigma \bar{\sigma}}^{\alpha \bar{\alpha}} (i\omega_m) {\tilde{\mathcal{F}^\dagger}}_{\bm{k},\sigma \bar{\sigma}}^{\alpha \bar{\alpha}} 
(i\omega_m) \right].
\end{align}
From above, the electromagnetic kernels in the static limit are given as follows.
\begin{align}
K_{\mu\nu}^p ({\bm q},0) &= +{\tilde K}_{\mu\nu}+ \delta {\tilde K}_{\mu\nu},\\
K_{\mu\nu}^d ({\bm q},0) &= -{\tilde K}_{\mu\nu}+ \delta {\tilde K}_{\mu\nu},
\end{align}
where ${\tilde K}_{\mu\nu}$ and $\delta {\tilde K}_{\mu\nu}$ are defined as follows.
\begin{align}
{\tilde K}_{\mu\nu} &= \delta_{\mu\nu} e^2 \int \frac{d{\bm k}}{\left( 2\pi \right)^3}  \frac{1}{\beta} \sum_{n=-\infty}^\infty \sum_{\al ,\sg}
v_{{\bm k}\al}^\mu v_{{\bm k}\al}^\nu \left( \tilde{\mathcal{G}^{ }}_{{\bm k},\sg \sg}^{\al \al} (i\omega_n) \right)^2 , \\
\delta {\tilde K}_{\mu \nu} &= -\delta_{\mu \nu} e^2 \int \frac{d{\bm k}}{\left( 2\pi \right)^3}  \frac{1}{\beta} \sum_{n=-\infty}^\infty \sum_{\al,\sg}
v_{{\bm k}\al}^\mu v_{-{\bm k}\bar{\al}}^\nu {\tilde{\mathcal{F}^{ }}}_{{\bm k},\sg \bar{\sg}}^{\al \bar{\al}} (i\omega_n) {\tilde{\mathcal{F}^\dagger}}_{{\bm k},\sg \bar{\sg}}^{\al \bar{\al}} (i\omega_n).
\end{align}
The total electromagnetic kernel is given by $K_{\mu\nu} = 2 \delta {\tilde K}_{\mu\nu}$. $\delta {\tilde K}_{\mu\nu}$ can be obtained as follows. First, we perform the summation for Matsubara frequencies of the anomalous Green's function.
\begin{align}
\frac{1}{\beta} \sum_{n=-\infty}^\infty {\tilde{\mathcal{F}^{ }}}_{{\bm k},\sg \bar{\sg}}^{\al \bar{\al}} (i\omega_n) {\tilde{\mathcal{F}^\dagger}}_{{\bm k},\sg \bar{\sg}}^{\al \bar{\al}} (i\omega_n)
&= \frac{1}{\beta} \sum_{n=-\infty}^\infty \sum_{\al'} \frac{\left| V_1 \right|^2 \left| V_2 \right|^2}{E_c^2}
\left( \frac{\theta \left( k_c - \left| {\bm k} - {\bm K}_{\al'} \right| \right)}{\left( i \omega_n - s_\al s_{\al'} E_{{\bm k}\al'+} \right)\left( i \omega_n - s_\al s_{\al'} E_{{\bm k}\al'-} \right)} \right)^2 \nonumber \\
&=\sum_{\al'} \frac{2 \left| V_1 \right|^2 \left| V_2 \right|^2}{E_c^2} \frac{\theta \left( k_c - \left| {\bm k}-{\bm K}_{\al'} \right| \right)}{( \xi_{{\bm k}\al'}^2 + 4 \left| V_{\al'} \right|^2 )^\frac{3}{2}}.
\end{align}

Therefore, $\delta {\tilde K}_{\mu\nu}$ is given as,
\begin{align}
\delta {\tilde K}_{\mu \nu} &=e^2 \sum_\al \int_{{\bm k}\in \mathcal{K}_\al} \frac{d{\bm k}}{\left( 2\pi \right)^3}  \frac{ k_\mu k_\nu}{m_\al {\tilde m}_\al} \frac{8 \left| V_1 \right|^2 \left| V_2 \right|^2}{E_c^2} \frac{1}{(\xi_{{\bm k}\al}^2 + 4 \left| V_\al \right|^2 )^{\frac{3}{2}}} \nonumber \\
&= \delta_{\mu\nu} \sum_\al \frac{2 \left| V_1 \right|^2 \left| V_2 \right|^2}{\left| V_\al \right|^2 E_c^2} \frac{n_\al e^2}{\left| {\tilde m}_\al \right|},
\end{align}
where we have used the relation $n_\al = (3/4) \rho_\al (0) |\mu_\al|$. Therefore, we can obtain the electromagnetic kernel represented in Eq.~(15).

%
